\documentclass[useAMS]{mn2e_modmargin}
\usepackage{amsmath}
\usepackage{url}
\usepackage{amsfonts}
\usepackage{amsbsy}
\usepackage{subfigure}
\usepackage{verbatim}
\usepackage{amssymb}
\usepackage{amsbsy}
\usepackage{graphicx}

\newcommand{\B}{\ensuremath{\mathbf{B}}}

\renewcommand{\u}{\ensuremath{\mathbf{u}}}
\renewcommand{\d}{\ensuremath{\partial}}

\newcommand{\ez}{\ensuremath{\mathbf{e}_{z}}}
\newcommand{\ephi}{\ensuremath{\mathbf{e}_{\phi}}}

\title[The MRI in cylindrical disk models]{Local and global aspects of the
linear MRI in accretion disks}
\author[Latter, Fromang \& Faure]{Henrik N. Latter$^{1}$\thanks{E-mail:
   hl278@cam.ac.uk},
   Sebastien Fromang$^{2}$,
   Julien Faure$^{2,3}$ \\
$^{1}$ DAMTP, University of Cambridge, CMS, Wilberforce Road,
Cambridge CB3 0WA, UK\\
$^{2}$ Laboratoire AIM, CEA/DSM-CNRS-Universit\'e Paris 7,
  Irfu/Service d'Astrophysique, CEA-Saclay, 91191 Gif-sur-Yvette,
  France\\
$^{3}$ Astronomy Unit, Queen Mary University of London, Mile End Road, London E1 4NS, UK}
\date{}

\begin{document}
\maketitle

\begin{abstract}

We revisit the linear MRI in a cylindrical model of an accretion disk and
uncover a number of attractive results
overlooked in previous treatments.
In particular, we elucidate the
connection between local axisymmetric modes and
global modes, and show that a local channel flow corresponds to the evanescent
part of a global mode. In addition, we find that the global problem reproduces 
the local dispersion relation without approximation, 
a result that helps explain the success the local analysis enjoys in predicting
global growth rates.
MRI 
channel flows are nonlinear solutions to the governing
equations in the local shearing box.
However, only 
a small subset of
MRI modes share the same property in global disk models,
providing further evidence
that the prominence of channels in
local boxes is artificial. Finally, we verify our results via direct numerical
simulations with the Godunov code RAMSES.

\end{abstract}


\section{Introduction}

The magnetorotational instability (MRI) remains the principal mechanism
facilitating turbulence, and consequently
mass accretion, in astrophysical disks. Since the seminal paper
of Balbus and Hawley (1991, hereafter BH91), numerical simulations
of MRI turbulence have become increasingly sophisticated and comprehensive;
now computations in global disk geometry are relatively common (e.g.\
Penna et al. 2010, Hawley et al.~2013, Parkin \& Bicknell 2013)
as is the inclusion of a panoply of physical effects 
(e.g.\ Jiang et al.~2013, Hirose et al.~2014, Lesur et al.~2014, Bai
2014). Despite this increase in complexity, BH91's
 local and
incompressible
linear theory continues to guide the
interpretation of the simulation results with surprising success (Flock
et al.~2010, Hawley et al.~2011, Okuzumi \& Hirose 2011).

When it first appeared, however, the analysis of BH91
provoked a debate concerning the local approach's validity when
describing phenomena in
the global geometry of an accretion disk. 
The emphasis of the debate focussed, in particular, on the small imaginary part
exhibited by the MRI growth
rates (neglected in the local analysis) as well as the boundary
conditions' impact on the magnitude of the growth
(also neglected; Knobloch 1992, Dubrulle \& Knobloch 1993). Ultimately, these concerns
were shown to be somewhat exaggerated.
Analyses in cylindrical models, vertically stratified boxes, and
general geometries indicated that 
the stability criterion was unaltered
and the local growth rates good approximations in most instances
(Kumar et al.~1994, Papaloizou \&  Szuszkiewicz 1994, Gammie \& Balbus
1994, Curry et al.~1994,
Curry \& Pudritz 1995, Terquem \& Papaloizou 1996, Ogilvie 1998).

One issue that remains undeveloped is the question of how
the BH91 local modes actually manifest in global disks. 
In other words: how do the
local and global formalisms join up? This is especially important for
the MRI channel modes, the fastest growing, because in the local
analysis they exhibit no radial variation at all. This implies that their
radial variation is in fact global. What might their radial structure be --- and does it
matter? A second
issue, also unexplored, is the nature of the modes that
participate in simulations of global MHD turbulence. Simulated modes are
never strictly local because of resolution constraints. We may ask
how well such modes can be understood via the local approach. We may
also want to construct a taxonomy of global modes and
determine how they control the
turbulence and other global aspects of the disk evolution (zonal
flows, winds, magnetic flux diffusion, etc). 

In this paper we make explicit the connection between the local and
global MRI, by revisiting the incompressible axisymmetric cylindrical
disk models employed by Dubrulle \& Knobloch (1993),  Coleman et
al.~(1994) and Curry et
al.~(1994). In so doing we uncover a number of attractive and
overlooked results. First, we show that the global problem
reproduces the (discretised)
local dispersion relation \emph{without approximation}.
The boundary conditions only determine the discretisation, they do not
generally influence the shape of the dispersion relation. 
Second, global MRI modes extend over a limited range of disk
radii before decaying in an outer evanescent region in which the
differential rotation is too weak.
We make clear
that local channel modes can be identified with these evanescent
portions,
 while local radially varying modes can be
identified with parts of the \emph{same} modes at smaller radii.
Third, convenient  analytic approximations to the global growth rates
 are available
via a matched WKBJ procedure. Fourth, in general, global MRI modes do
not possess the nonlinear property of familiar local channel flows
(Goodman \& Xu 1994). Only 
 modes localised to the inner edge of the
disk remain acceptable solutions to the
full equations when possessing nonlinear amplitudes. 
These results are verified
numerically with a small set of simulations using the Godunov code,
RAMSES (Teyssier 2002, Fromang et al.~2006). Finally, we discuss the
issues and questions they raise, and future work that could
begin to address them.

\section{Linear stability analysis}
\label{sec:linear_stability}

Within this paper we consider an annular slice through the midplane of
the disk. The slice's vertical thickness is taken to be much less
than the disk's vertical scale height.
The background vertical
structure is hence neglected, and the model is
quasi-global: `local' in the vertical and
`global' in the radial. 
For convenience we also dispense with the disk's radial
structure
and treat the ionised fluid as incompressible. 
The resulting formalism remains mathematically tractable while
retaining
important global effects (boundary
conditions and curvature).

The first linear calculations in this set-up were undertaken by
Velikhov (1959) and Chandrasekhar (1961), but it was not until the
1990s that accretion disks were explicitly modelled in this way
(Knobloch 1992, Dubrulle \& Knobloch 1993, Kumar et al.~1994, Curry et
al.~1994), the latter work generally confirming the local theory of
BH91. More recently, Kersal\'e et al.~(2004) generalised the set-up to
permit a radial flow of material through the inner boundary; and, while
the classical MRI is recovered, spurious `wall modes' are generated by
the inner boundary condition. These modes control to some
extent the ensuing nonlinear dynamics (Kersal\'e et al.~2006).
Finally, we note the work of Rosin \& Mestel (2012)
which included the Braginskii stress, and hence could
describe the onset of
instability in the
weakly collisional plasma of the Galactic disk. 

In this section we first exhibit the main equations and
repeat the classical local axisymmetric stability analysis for
reference. 
Next the linearised equations in cylindrical geometry are reduced to a
single second-order Sturm-Liouville equation, through which we display the
similarities with the local problem. Numerical and analytic solutions are
then derived.

\subsection{Governing equations}
\label{sec:gov_eq}
We work with the equations of ideal incompressible MHD,
\begin{align}
\d_t \u + \u\cdot\nabla\u &= -\nabla\Phi - \frac{1}{\rho}\nabla P_t +
\frac{1}{4\pi\rho}\B\cdot\nabla\B, \\
\d_t \B + \u\cdot\nabla\B &= \B\cdot\nabla \u, \\
 \nabla\cdot\u &=0 \\
 \nabla\cdot\B &=0.
\end{align}
Here velocity and magnetic field are denoted by $\u$ and $\B$
respectively, $\rho$ is the constant density,
 $\Phi$ is the gravitational potential of the central
object, and $P_t$ is the combined gas and magnetic pressure. 
The accretion disk is a circular annulus, with gas
inhabiting cylindrical radii between $r=r_0$ and $r=r_1$ with
$r_0\ll r_1$. The vertical
extent of the domain we set to positive and negative infinity. This
setup describes motions that possess
short vertical lengthscales (shorter than the scale height) and
up to long radial lengthscales ($\gtrsim r_0$). 

\subsubsection{Equilibrium}

The governing equations admit the steady equilibrium solution:
\begin{align}
\u= r\Omega(r)\,\ephi, \quad \B= B_0\,\ez, \quad P_t= P_t(r),
\end{align}
The rotation frequency is a power law,
$$ \Omega = \Omega_0\left(\frac{r}{r_0}\right)^{-q}.$$
If $q=3/2$ the disk is Keplerian and the background pressure is
constant (no pressure gradient is required for radial force balance).
When $q\neq 3/2$ the pressure gradient is non-zero, but its details
are not
important in what follows (see Curry et al.~1994 for further specifics). 

\subsubsection{Perturbations}

The equilibrium is disturbed by axisymmetric modes of the type 
$\propto F(r) e^{\text{i} k_z z + st}$,
where $F$ is an $r$-dependent
perturbation amplitude, $k_z$ is the (real) vertical wavenumber, and $s$ is the
(potentially complex) growth rate. The assumption of vertical
locality means that $k_z r_0 \gg 1$.

The ensuing linearised equations are
\begin{align} \label{l1}
s u_r' - 2 \Omega u_\phi' &= -\d_r h' + i k_z v_A^2 b_r', \\
s u_\phi' + (2-q)\Omega u_r' &= i k_z v_A^2 b_\phi', \\
s u_z' &= - i k_z h' + i k_z v_A^2 b_z', \\
 s b_r' &= i k_z u_r', \\
 s b_\phi' &= i k_z u_\phi' - q\Omega b_r', \\
s b_z' &= i k_z u_z',\\
(1/r)\d_r(r u_r') + i k_z u_z' &=0. \label{l7}
\end{align}
Here a prime indicates the perturbation, while
$\mathbf{b}'= B_0\B'$ and is hence dimensionless. The Alfven speed is
defined by $v_A^2=B_0^2/(4\pi\rho)$ and the enthalpy by $h'=P_t'/\rho$.  
Note that these perturbations automatically satisfy the solenoidal
condition on $\mathbf{b}'$. 

To complete the problem we must supply
two boundary conditions,
applied to the radial velocity $u_r'$. The boundaries can be treated
as hard walls or as stress free (Dubrulle \& Knobloch 1993), 
in which case either
$u_r'$ or $\d_r u_r'$ is zero at $r=r_0$ and $r=r_1$, or they may be
treated as free
surfaces (Curry et al.~1994), in which case a linear combination of
$u_r'$ and $\d_r u_r'$ is zero.
As we will see later, it is not terribly important which we choose,
only that the conditions are homogeneous.

Note that the Alfven frequency $k_z v_A$ associated
with each mode is constant throughout
the disk. In contrast, the orbital frequency $\Omega(r)$ decreases with
radius. Consequently, at sufficiently large radius 
magnetic tension
dominates and
the conditions for MRI become unfavourable. 
We expect a growing MRI mode of given $k_z$ to avoid such radii and instead emerge
closer to the central body. If, however, the
background magnetic field decays with radius faster than $\Omega$,
this need not be the case.

\subsection{Local axisymmetric dispersion relation}

In order to examine local modes, we 
choose a point $r_*$ and examine the behaviour of the gas 
in its immediate vicinity. For axisymmetric disturbances the orbital
frequency $\Omega$ can be regarded as constant in Eqs
\eqref{l1}-\eqref{l7}, and if their radial variation is small-scale
then the cylindrical term in Eq.~\eqref{l7} may be dropped. 
This permits us to decompose the disturbances in Fourier modes 
 $\propto\text{e}^{\text{i}k_x r}$, and
the local MRI dispersion relation for such modes is
straightforward to derive:
\begin{align}\label{kxlocal}
s^4 + [2 k_z^2 v_A^2 + 2(2-q)\epsilon^2 \Omega_*^2]s^2 + k_z^2 v_A^2\left(k_z^2 v_A^2 -
2q\epsilon^2\Omega_*^2 \right)=0.
\end{align}
Here $\Omega_*=\Omega(r_*)$, the rotation rate at the radius in which
we're interested, and
\begin{align} \label{localdisp}
\epsilon = \frac{1}{(1+ k_x^2/k_z^2)^{1/2}}.
\end{align}

In Eq.~\eqref{kxlocal} the radial wavenumber $k_x$
appears solely in $\epsilon$ and then only in the ratio
$k_x/k_z$. Note also that instances of $\epsilon$ 
occur exclusively as factors of
$\Omega_*$. 
The familiar channel flows are obtained when $k_x/k_z \to 0$, with the modes
exhibiting little (to no) relative radial variation. In this case,
$\epsilon\to 1$ and the fastest growth rate is achieved,
$q\Omega_*/2$. 
Because $\epsilon \Omega_*$ sets the timescale of the growth rate, modes
with non-zero $k_x$ grow slower, as they possess $\epsilon< 1$. We call radially
varying modes `radial modes' to distinguish them from channel modes.

The
local dispersion relation \eqref{kxlocal} exhibits an interesting
connection between radial \emph{variation}, on one hand, and radial
\emph{location}, on the other.   
By varying $k_x$ (and keeping $k_z$ fixed) we can `rescale'
the MRI timescale $\epsilon\Omega_*$, because the factor $\epsilon$ depends on
$k_x$. A striking consequence is that
a radial mode of a given $\epsilon$ located at one radius $r_i$, 
associated with an orbital frequency of $\Omega(r_i)= \Omega_*$,
possesses the \emph{same} growth rate 
as a channel mode located at a different larger radius $r_o>r_i$, 
associated with an orbital frequency of
$\Omega(r_o)=\epsilon\Omega_*$. 
As a consequence, we can `map' modes of
different $k_x$, but different radial locations, onto one another. 

Physically, one can understand the radial modes by returning to the
classical cartoon of the MRI mechanism (for e.g.\ BH91). 
The upper panel of Fig.~1 
illustrates a channel mode, with the blue circles representing fluid
blobs and the black lines indicating the magnetic field (initially
vertical). Vertically varying radial perturbations lead to the
periodic stretching of the magnetic field and, consequently, to angular
momentum exchange between the tethered blobs. The instability proceeds
from the counterstreaming flows that
ensue.

\begin{figure}
\scalebox{0.425}{\includegraphics{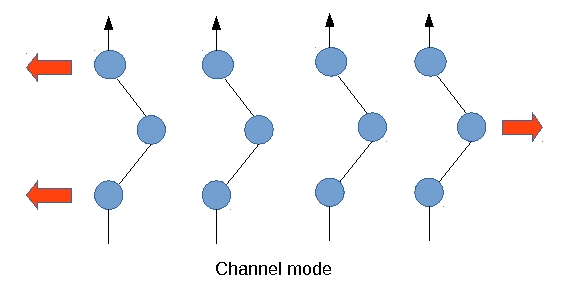}}
\scalebox{0.425}{\includegraphics{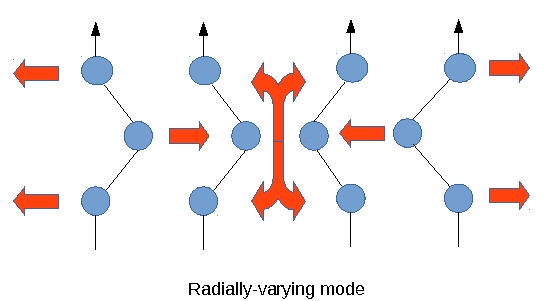}}
 \caption{Simple illustrations of a channel mode (upper panel) and a radial
   mode (lower panel). Blue circles represent fluid blobs, the
 black arrows indicate the initially vertical magnetic field. Red
arrows show the direction of the fluid motion.}
\end{figure}

However, if the mode possesses radial structure there will exist nodes
in radius
where the streams converge and diverge, illustrated in the lower panel
of Fig.~1. At these radii the mode exhibits vertical deflection and
pressure perturbations which are not present in pure channel flow.
Needless to say, the development of vertical circulations impedes the
instability mechanism exemplified in pure channel flow. 
During their vertical excursions fluid blobs stop
extracting free energy because they 
are no longer exchanging angular momentum effectively.

\subsection{Global eigenvalue problem}

We now return to the linearised global equations
\eqref{l1}-\eqref{l7} which can be reworked
into a single second-order equation for the variable $U= r^{1/2}
u_r'$. Rescaling space by $r_0$ we find
\begin{equation} \label{Und}
\frac{d^2 U}{dr^2} -\left( \frac{3}{4r^2} +k_z^2 \right)U= -\frac{k_z^2}{r^{2q}}\varepsilon^{-2}\,U,
\end{equation}
which should be solved on the domain $r\in [1,\,r_1/r_0]$. Here 
the various parameters appearing in the
problem, in addition to the unknown growth rate $s$, have been
packaged into the convenient quantity $\varepsilon$ defined via
\begin{equation}\label{lambda}
\varepsilon^{-2} = 2\Omega_0^2\,\frac{(q-2)s^2+q k_z^2
  v_A^2}{(s^2+k^2_z v_A^2)^2}.
\end{equation}
In the above, $\Omega_0=\Omega(r_0)$, the rotation rate at the inner
boundary.

With homogeneous boundary conditions (such as impenetrable walls
or a free surface), Eq.~\eqref{Und} is in Sturm-Liouville form. 
The weight function is $-k_z^2\,r^{-2q}$ and the eigenvalue is
$\varepsilon^{-2}$. Because the problem is Sturm-Liouville, 
we are assured of a discrete set of real
eigenvalues $\varepsilon_n^{-2}$ 
which we order so that 
$\varepsilon_0 > \varepsilon_1 > \varepsilon_2 > \dots $. 
Once these are computed
the associated growth rates can be obtained. Hence the problem is
broken down into two steps: first calculate $U$ and $\varepsilon$ from
\eqref{Und}, then
calculate the growth rate $s$ from \eqref{lambda}.
Note that the eigenfunction structure $U$
depends only on $k_z$; it is oblivious to the magnetic field strength
$v_A$ unless it appears in the boundary conditions.

\subsubsection{Global dispersion relation}

Equation \eqref{lambda}, which defines the eigenvalue $\varepsilon$,
 can be reworked into
\begin{equation}\label{dispy}
s^4 + [2 k^2_z v_A^2 + 2(2-q)\varepsilon_n^2\Omega^2_0]s^2 + k^2_z v_A^2\left(k^2_z v_A^2 -
2q\varepsilon_n^2\Omega_0^2 \right)=0,
\end{equation}
which is almost identical to
Eq.~\eqref{kxlocal}! 
Remarkably, the axisymmetric global problem yields a variant
of the local axisymmetric dispersion relation. The basic structure of
the local problem is retained independently of the global specifics: 
curvature and boundary conditions.

There are two differences. Instead of the
function $\epsilon$, which depends smoothly on the continuous radial wavenumber
$k_x$, the global problem possesses the discrete function
$\varepsilon_n$ which depends on the radial quantum number $n$. Unlike
the local problem, we do not know how $\varepsilon_n$ depends on $n$
a priori; this information must be extracted from the
differential equation \eqref{Und} and depends on $k_z$, $q$, and the
boundary conditions. 
However, it is easy to show that
$\varepsilon_n<1$ (see Appendix A). Hence we order the eigenvalues as
$1> \varepsilon_0 > \varepsilon_1 > \varepsilon_2 > \dots > 0$.
The second difference is that $\Omega$ is set to
$\Omega_0$, the orbital frequency at the inner radius of the disk. This
fixes the minimum timescale over which appreciable growth can happen. 

It should be emphasised that, in contrast to the claims of previous
authors,
the only role the boundary conditions
play is to help set the discretisation of the dispersion relation
\eqref{dispy}. 
It does not alter the shape of the relation itself. That fundamental
shape is determined by the \emph{local} physics, as first explained in
BH91. Moreover, as we show later, the larger $r_0k_z$ the more
closely spaced the $\varepsilon_n$ and hence the less
important the discretisation.
 
\subsubsection{Stability criterion}

Next we rearrange Eq.~\eqref{Und} so that it resembles a Schr\"odinger
equation,
\begin{equation} \label{Und2}
\frac{d^2 U}{dr^2} +k_z^2\,f(r,\varepsilon)\,U= 0,
\end{equation}
where
\begin{equation} \label{f}
f(r,\varepsilon)= \frac{1}{r^{2q}}\varepsilon^{-2}
 - \frac{3}{4r^2k_z^2} - 1.
\end{equation}
The problem now resembles that of a particle in a potential well, with
the potential
proportional to $f$. It is easy to show that $f$ has an extremum at 
\begin{equation}
r= \left(\tfrac{4}{3}qk^2/\varepsilon^2\right)^{1/[2(q-1)]}.
\end{equation}
In order for there to be trapped waves, i.e.\ normal modes, the
extremum must lie inside the disk. For simplicity we let its outer
boundary go to infinity, then this condition, combined with \eqref{dispy}, gives us
a general instability criterion in terms of the vertical field
\begin{equation} \label{stabcrite}
M_A > \frac{1}{q}\sqrt{\frac{3}{8}},
\end{equation}
where $M_A$ is the Alfv\'enic Mach number of the background state,
defined to be equal to $\Omega_0
r_0/v_A$. The criterion \eqref{stabcrite} gives an upper
bound on the magnetic field threading the disk. Because in most contexts 
the field is assumed to be subthermal and the disk assumed thin,
we have $M_A \gg 1$ and the
criterion is automatically satisfied. A far more restrictive condition
on the magnetic field issues from the disk's vertical
thickness (see BH91 and Gammie \& Balbus 1994).

\subsubsection{Turning points and global and local modes}

At sufficiently large $r$ the function $f$ will be dominated by the
last term in Eq.~\eqref{f}. In this region $U \sim \text{e}^{-k_z r}$ and
the mode is evanescent. Physically this makes sense: sufficiently
far out in the disk the differential rotation is too weak to compete
with magnetic tension and the MRI mechanism is suppressed. Active MRI
modes will shun such a region and instead localise at smaller radii
where the conditions for instability are more favourable. 
In Fig.~2 we present a representative eigenfunction illustrating this
morphology.

\begin{figure}
\scalebox{0.425}{\includegraphics{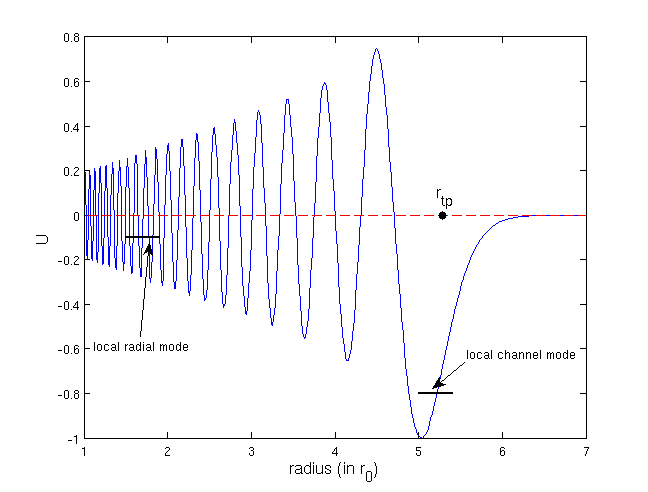}}
 \caption{A higher order global eigenfunction of
   \eqref{Und}. The turning point $r_\text{tp}$ is indicated as are
   regions of the global mode that we
   identify with local channel and radial modes. Note also the monotonic
   decrease in the local radial wavenumber $k_r$ with radius.}
\end{figure}

The boundary
between the evanescent region and the `MRI region' is given by the
turning point of Eq.~\eqref{Und2}, i.e.\ the radius at which $f=0$,
denoted by $r_\text{tp}$. 
For general $q$ and $k_z$ this must be determined
numerically. However, in the plausible limit of $k_z$ large,
\begin{equation}
r_\text{tp} \approx \varepsilon^{-1/q}.
\end{equation}
In fact, this radius corresponds to an orbital frequency of $\Omega =
\varepsilon\Omega_0$. 

Combining this last piece of information with the local and global
dispersion relations reveals an
interesting correspondence. A global mode, associated with a given
$\varepsilon_n$, shares the same dispersion relation \eqref{dispy} as that
of a local channel
mode situated at the global mode's turning point \eqref{kxlocal}.
This is because the local orbital frequency $\Omega_*$ 
at this point is precisely $\varepsilon_n\Omega_0$. 
Consequently, we are invited to identify a local channel mode as the
small section of a much larger global mode --- the section
near evanescence. Indeed, we may define a (radius dependent) 
radial wavenumber via $k_r \approx k_z\,f^{1/2}$, and at the turning point
$k_r=f^{1/2}=0$, by definition, which is in accord with the channel
mode's lack of any radial structure.

Meanwhile, what of local radial modes? 
First, consider a radial mode located
at $r\approx r_0$, i.e.\
very near the inner boundary, and possessing a $k_x$ so that $\epsilon
\approx \varepsilon_n$, for the same $n$ as above. 
This particular radial mode now shares the same dispersion relation as
the previous channel mode, and hence the same dispersion relation as
the global mode, Eq.~\eqref{dispy}. Hence we may also
identify this
local radial mode as part of the same parent global mode ---
but the part of that mode
near the inner
boundary. Moreover, we can do exactly the same with other radial modes
of smaller $k_x$. But for each different $k_x$ we must change the
local mode's radial location so that the timescale $\epsilon\Omega$ is kept
constant. 

This explains the connection between radial
variation and radial location that appeared in the local dispersion
relation of Section 2.2. This connection 
arises because a global mode of given $n$ is constituted from
a set of local modes of differing
$k_x$, each at a different radial location.
In Fig.~2 this idea is sketched out.

It is worth stressing that the 
growth rate of the global mode is limited by the local physics at the
mode's periphery, at $r=r_\text{tp}$.
Though the mode extends over regions where the local growth
rate can be faster, the mode can only grow as fast as its outer `edge'. 
Put another way: the structure, as a whole, can only grow
as fast as its slowest component. This helps remove some of the
ambiguity when attributing local growth rates to
global simulations.
It is clear that at any given radius, growth is controlled by the mode
whose turning point falls at that radius.
No other mode that extends to this
location can grow faster (faster growers are localised to
radii closer in). Moreover, as we see from Figs 2 and 3, modes'
amplitudes are maximal near $r=r_\text{tp}$.
Indeed, numerical simulations verify this.
At any given radius, growth occurs at the
rate of the local channel mode (Hawley 2001), which of course is also
the rate of the global mode whose turning point
$r_\text{tp}$ falls there.

\subsection{Global solutions}
\subsubsection{Numerical solutions}
\label{sec:num_eigen}

Having sketched out the background details, we present in this
subsection 
a few numerical solutions to 
\eqref{Und} subject to the hard wall boundary conditions, $U=0$ at
$r=1,\,r_1/r_0$. 
Our domain is set to be $r \in [1,10]$. The only parameter that appears
explicitly in \eqref{Und} is the vertical wavenumber $k_z$. We let it equal
10 for our main results.
The numerical technique we use is a pseudo-spectral Chebyshev method
(Boyd 2002), which approximates the differential equation by a
matrix. Its eigenvalues may be obtained by the QZ algorithm
(Golub \& van Loan 1996). The boundary conditions are encoded in the
matrix via boundary bordering.

\begin{figure*}
\scalebox{0.35}{\includegraphics{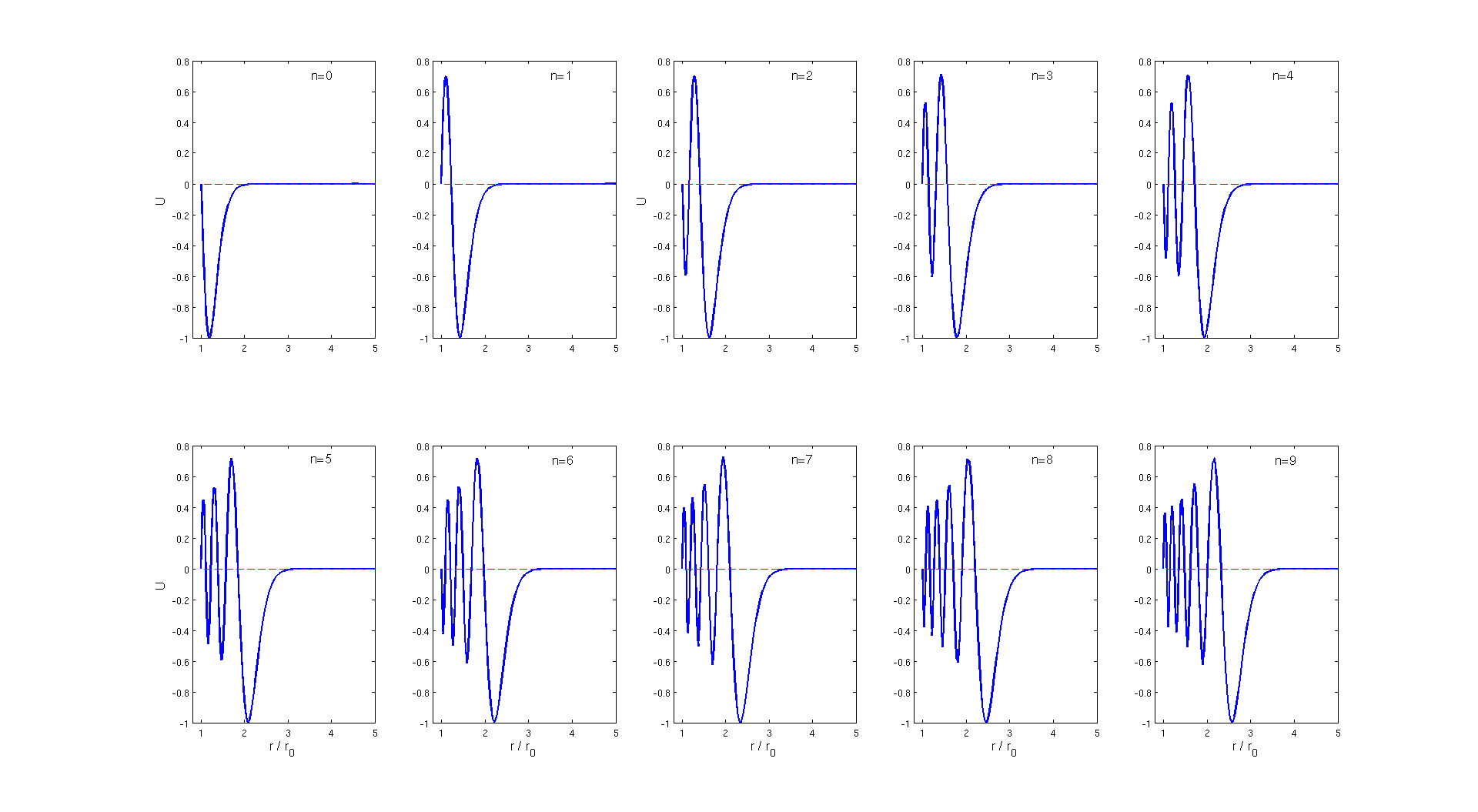}}
 \caption{The first ten MRI eigenfunctions when $k_z=10$. In
   each case, the modes are oscillatory within the region
   $1<r<r_\text{tp}$ and clearly evanescent outside these. Real parts
   are plotted; the imaginary parts are zero.}
\end{figure*}

In Fig.~3 appear the first ten eigenfunctions for $k_z=10$. The
associated $\varepsilon_n$ for the first four are 
$$ \{0.6413,\,0.4976,\,0.4169,\,0.3628 \},$$
with corresponding turning points $r_\text{tp}$
$$\{ 1.345,\,1.593,\,1.792,\,1.966 \},$$
where the mode transitions
to an evanescent wave. As anticipated, larger $n$ correspond
to modes that extend over more of the domain.
 
In order to compute explicit growth rates we need to specify the
strength of the vertical magnetic field, this can be measured in terms
of either the 
dimensionless combination 
$v_A (k_z/\Omega)$ or the Alfv\'enic Mach number $M_A$. 
Selecting the first option, we plot in Fig.~4
the growth rates of the first four modes as functions of $v_A$ and
fixed $k_z=10$. These are just four copies of the standard local MRI
relation.
The different curves correspond, not to differences in the modes' vertical
structure --- this is held fixed --- but to different radial
structures. The greater the radial quantum number $n$ the more radial
structure, but most importantly the greater $r_\text{tp}$ and hence
the slower the dynamical timescale. This explains why higher $n$ modes
grow slower.
It should be appreciated that the fastest growing mode need
not determine the evolution of the disk globally. The fastest growing
modes are localised at small radius. Larger radii will be
driven by the slower modes that extend to them. 

\begin{figure}
\scalebox{0.45}{\includegraphics{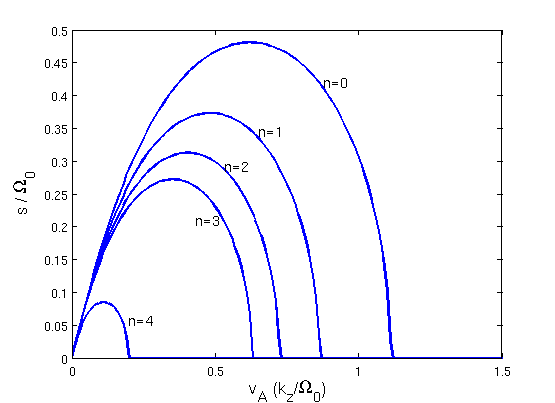}}
 \caption{MRI growth rates for five different radial quantum numbers and
 when $k_z=10$ but $v_A$, the strength of the background
 magnetic field, varies. Note that $v_A$ has been scaled using a fixed
 $k_z$; this is so the classical MRI dispersion curves are easier to see. }
\end{figure}

\begin{figure}
\scalebox{0.5}{\includegraphics{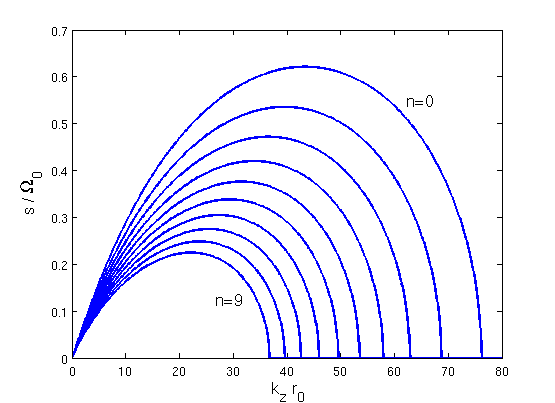}}
 \caption{MRI growth rates for the first 10 radial quantum number $n$
   when $v_A$ is fixed and $k_z$ is varied. We have set
   $M_A= r_0\Omega_0/v_A= 50$. }
\end{figure}

Lastly, we fix $v_A$ and see how the growth rates vary as a function
of $k_z$. In Fig.~5 we plot $s$ as a function of $k_z$,
 different $n$, and for a weak field fixed by $M_A =
50$. 
These curves are similar but not exactly
those of the standard MRI, seen in Fig.~4. That is, we cannot simply
rescale $k_z r_0$ by $k_z v_A/\Omega_0$ and recover the same relation. This
is because $\varepsilon_n=\varepsilon_n(k_z)$.

\subsubsection{Special solutions}

For certain special cases, Eq.~\eqref{Und} can be solved exactly.
When the disk exhibits a rotation profile of $q=1$, roughly similar to the Galaxy,
we find that 
\begin{equation} \label{Galactic}
U = r^{1/2}K_\nu(k_zr), \qquad \text{with} \qquad \nu= \sqrt{1-k_z^2/\varepsilon^2},
\end{equation}
and where $K_\nu(x)$ is the modified Bessel function of the second
kind. In obtaining \eqref{Galactic} we have taken the outer boundary
to infinity and then applied the outer boundary condition, $u'_r= r^{-1/2}U\to 0$. 
The eigenvalue equation for $\varepsilon$ is $K_\nu(k_z)=0$ in the case
of a hard inner wall. 
These solutions were explored first by Dubrulle \&
Knobloch (1993), and more recently by Rosin \& Mestel (2012).

In the limit of large $k_z$, approximations to the
eigenvalues can be obtained from
\begin{equation} \label{Galeigs}
\varepsilon_n \approx 1 + a_n\, 2^{-1/3}\,k_z^{-2/3},
\end{equation}
where $a_n$ is the $n$'th negative root of the Airy function
$\text{Ai}(x)$. A brief derivation of this expression is given in
Appendix A2 (see also Rosin \& Mestel 2012). 
Note that in this limit the spacing between neighbouring
eigenvalues becomes tiny, and thus the dispersion relation
\eqref{dispy} is effectively continuous: local physics dominates the
global mode.  

When the rotation profile is Keplerian, $q=3/2$, there is no general
analytic 
solution to Eq.~\eqref{Und}. However, in the limit of small $k_z$ we have 
\begin{equation}
U \approx r^{1/2} J_2\left(\frac{2k_z}{\varepsilon}r^{-1/2}\right),
\end{equation}
where $J_2(x)$ is the Bessel function of the first kind and second order. The
ensuing eigenvalues are
\begin{equation}
\varepsilon_n = 2k_z/b_n,
\end{equation}
where $b_n$ is the $n$'th root of $J_2(x)$.
Of course, the
limit of small $k_z$ is not the most appropriate for our cylindrical
disk model, as it indicates the mode is global in the $z$-direction.
The result nevertheless offers a useful test of the numerical
solver in Section 2.4.1.

\subsubsection{WKBJ solutions}

In this section we present asymptotic solutions to
\eqref{Und} in the limit of large vertical
wavenumber $k_z$ and for general $q$. 
As explained earlier, this is a natural limit because we expect $k_zr_0$ to be
large. 

The solution
is oscillatory between $r=1$ and $r=\varepsilon^{-1/q}=r_\text{tp}$, and it is
evanescent for $r> r_\text{tp}$.
Within the former region we employ the standard WKBJ
ansatz:
\begin{equation} \label{wkbjalltheway}
 U = |f|^{-1/4}\,\cos\left[k_z\int_{r_\text{tp}}^r |f|^{1/2} dr + \pi/4\right].
\end{equation}
The phase shift of $\pi/4$ comes from matching across the turning
point (see for e.g.\ Riley et al.~2006). 
To obtain the eigenvalue equation for $\varepsilon$ we next impose the 
hard-wall boundary condition
 at $r=1$, i.e. $U=0$. This leads to
\begin{equation}\label{disp}
\int_1^{\varepsilon^{-1/q}}
\left(\frac{1}{\varepsilon^2 r^{2q}}-1\right)^{1/2}\,dr = \frac{\pi}{k_z}\left(n+\frac{3}{4}\right),
\end{equation} 
where $n$ is the radial quantum number and in which we have set
$f(r)= (\varepsilon r^q)^{-2} -1$ to leading order in $k_z$. Notice,
 that for large $k_z$ neighbouring $n$ modes posses
eigenvalue equations that differ only marginally on the right hand side. As a
consequence, the values of the corresponding $\varepsilon_n$ are extremely close to one
another, as in Eq.~\eqref{Galeigs}.

The integral in \eqref{disp} can be written in terms
of special functions by introducing the integration variable
$\xi=\varepsilon^{2}r^{2q}$.
 The eigenvalue equation then becomes
\begin{equation}
B\left(\tfrac{1}{2q}-\tfrac{1}{2},\,\tfrac{3}{2}\right)
 -
 B_{\varepsilon^2}\left(\tfrac{1}{2q}-\tfrac{1}{2},\,\tfrac{3}{2}\right)
= \frac{2\pi\,q}{k_z}\,\varepsilon^{1/q}\left(n+\tfrac{3}{4}\right),
\end{equation}
where $B(x,y)$ and $B_\mu(x,y)$ are the complete and incomplete beta
functions (Abramowitz \& Stegun 1964). Though it is relatively
straightforward to numerically solve this nonlinear equation, we have
the following convenient approximations. The fastest growing modes
possess $\varepsilon$ near 1, which may be approximated by
\begin{equation} \label{near1}
\varepsilon_n \approx 1 - \tfrac{1}{2}\left[3\pi q\left(n+\tfrac{3}{4}\right)\right]^{2/3}\,k_z^{-2/3}.
\end{equation}
Note the similarity to Eq.~\eqref{Galeigs}\footnote{Incidentally, when
  $q=1$ this equivalence
  provides a neat analytic approximation to the zeros of Ai($x$). 
For a more direct derivation see Fabijonas and Olver (1999).}.
For small and intermediate $\varepsilon$ a different expansion of the
beta functions yields
\begin{equation} \label{approx}
A\varepsilon^{-1/q} + \frac{1}{q-1}\varepsilon^{-1} + \frac{1}{2(1+q)}\varepsilon +\frac{\pi}{k_z}\left(n+\frac{3}{4}\right)=0,
\end{equation}
where $A$ is the number
\begin{equation} \label{AA}
A=
\frac{\sqrt{\pi}\,\Gamma\left[\frac{1}{2q}+\frac{1}{2}\right]}{2\Gamma
\left[\frac{1}{2q}\right]},
\end{equation}
 and $\Gamma[x]$ is the gamma function. A Keplerian rotation law
reduces Eq.~\eqref{approx} to a quintic polynomial equation.

 For comparison with the computations in Section 2.4.1,
 the WKBJ
eigenvalues $\varepsilon$ are
$$ \{0.645,\, 0.499,\,0.418,\,0.363\},$$
which agree reasonably well. 
The agreement improves as $k_z$ increases. Values of
$\varepsilon$ gathered from either \eqref{near1} or \eqref{approx} may
be input into the dispersion relation \eqref{dispy} and the
growth rates computed without the need to numerically solve the ODE,
which is the main benefit of the WKBJ approach.

\section{Are global modes nonlinear solutions?}

In classical and vertically stratified shearing boxes, channel flows are
nonlinear solutions in the incompressible and anelastic regimes,
respectively (Goodman \& Xu 1994, Latter et al.~2010). It is then natural
to ask if global modes in cylindrical geometry possess an analogous
property. Indeed, the initial stages of some
nonlinear simulations (e.g.\ Hawley 2001) exhibit strong counterstreaming
flows in the inner parts of the disk that suggest this might be the
case. In this section, however, we show that only a small subset of
the linear global modes can be said to be approximate nonlinear
solutions, and furthermore these are localised very close to the inner
boundary. 
 
The key feature of local channel flow is the
strong separation between their radial and vertical
lengthscales. We hence introduce the small parameter 
\begin{equation} 
 \delta = 1/(k_z \lambda_r),
\end{equation}
where $\lambda_r$ is the modes' characteristic radial
lengthscale. From Eqs \eqref{l1}-\eqref{l7} the following scalings can
be derived
\begin{align*}
&u_z' \sim \delta\,u_r', \qquad u_\phi'\sim u_r' \\
&b_z' \sim \delta\,b_r', \qquad b_\phi'\sim b_r',
\end{align*}
in addition to $h' \sim \delta^2 (\lambda_r\Omega) u_r'$. 
Thus in the limit of small $\delta$ the morphology of global modes resembles
local channel flows: vertical velocity and magnetic field is minimised, as is
the pressure perturbation.
Perfect
channel flows cannot be obtained, however, because of the cylindrical terms in
conjunction with incompressibility and the solenoidal condition. 
Some (small) radial variation in
the mode structure must give rise to (small) vertical motion or field.
In the local approximation $k_z$ and $\lambda_r$ can be
chosen independently and $\delta$ can be made as small as required. 
This is not the case for global modes, however, and generally
$\lambda_r=\lambda_r(k_z)$. We defer estimates of the magnitude of
$\delta$
to later in this section; for the moment we assume that $\delta$ can
be made sufficiently small.

Let us now inspect the size of the non-linearities associated with the modal
solutions. We want to estimate at what point the nonlinearities become important.
Assuming the above scalings, 
the ratio of the nonlinear advective term in the
radial component of the momentum equation 
to the linear terms is
\begin{align}
\frac{(\u'\cdot\nabla\u')_r}{s u'_r}& 
 \sim \frac{u_r'\d_r u_r' + u_z' \text{i} k_z
  u_r' - u_\phi'^2/r}{\Omega u_r'}, \\
& \sim \frac{u_r'}{\lambda_r\Omega}.
\end{align}
To progress further, we assume that the mode began growing at a
fraction $a$ of the background Alfven speed $v_A$, and thus $|\u'|
\propto a\, \text{e}^{st} v_A$ (see Goodman \& Xu 1994). Placing this in
the above scaling gives 
\begin{align}
\frac{(\u'\cdot\nabla\u')_r}{s u'_r} & \sim
\frac{v_A}{\lambda_r\Omega} a\,e^{st} \\
& \sim \delta\,a\, \text{e}^{st}, \label{scaling}
\end{align}
where we have assumed that $v_A k_z/\Omega \lesssim 1$, in order for the
MRI to work. 
A similar argument gives the same scaling for the other
components of the advective term, as well as the $\B\cdot\nabla\B$ and
mixed terms. For instance, using \eqref{l1}-\eqref{l7},
\begin{align}
\frac{(\B'\cdot\nabla\B')_r/(4\pi\rho)}{s u'_r}& \sim
    v_A^2  \frac{b_r'\d_r b_r' + b_z' \text{i} k_z
  b_r' - b_\phi'^2/r}{\Omega u_r'}, \\
& \sim \frac{v_A^2 k^2}{\Omega^2}\frac{u_r'}{\Omega\lambda_r}, \\
& \sim \frac{u_r'}{\Omega\lambda_r},
\end{align}
and the same
scaling as \eqref{scaling} is recovered.

In all cases the quadratic nonlinearities go as $\delta a
\,\text{e}^{st}$ relative to the linear terms. This immediately gives us a
condition for when they are important and when the MRI deviates from the
linear channel structure. We set $\delta a\,
\text{e}^{st}$ to 1, and compute the time at when this occurs. The critical
time $t_1$ is 
\begin{align}
t_1 \sim \frac{1}{s} \ln\left(\frac{1}{\delta a}\right).
\end{align}
It follows that
the nonlinearities only become
important once the mode amplitudes have grown a factor $\delta^{-1}$ greater
than the background field. Thus the smaller $\delta$ the more dramatic
the amplification of the linear cylindrical modes, and the deeper they
penetrate the nonlinear regime. 

In fact, for most parameter choices we find only $\delta \lesssim
1$. This is because the larger we take the value of $k_z$, the smaller 
the corresponding $\lambda_r$, and as a result $\delta$ need not
be small. Most modes, consequently, cannot be said to possess the
nonlinear property of MRI channel flows: once their amplitudes reach
that of the background field, quadratic nonlinearities become
important and the modes break down.

However, for small $n$ and exceedingly large $k_z$ the radial and
vertical scales can be separated and $\delta \ll 1$ is achievable. 
The scaling of
$\delta$ with large $k_z$ is straightforward to obtain. First
assume that for small $n$ the radial scale of the mode is roughly
equal to the distance between $r=1$ and its turning point
$r=r_\text{tp}$. Thus $\lambda_r \approx \varepsilon_n^{-1/q}-1$. 
Equation \eqref{near1} provides an estimate of $\varepsilon_n$ in the
limit of large $k_z$, and we get $\lambda_r \sim k_z^{-2/3}$. This
returns 
\begin{equation} \label{epscale}
\delta \sim k_z^{-1/3},
\end{equation}
a rather weak scaling. As a consequence, small values of $\delta$ require
exceptionally large values of $k_z$. Indeed, only modes with $k_z>10^{3}$
possess anything resembling the nonlinear property seen
in local boxes. Moreover the radial extent of such modes scale like
$k_z^{-2/3}$ and so they are effectively localised to the inner edge of
the disk. As a consequence, 
the vast majority of an accretion disk will never
 experience
the nonlinear property of the linear MRI modes.

We finish by discussing alternative mechanisms that disrupt those few
global modes that do possess an approximate form of the nonlinear
property. Compressibility is perhaps the primary
mechanism in local boxes (Latter et al.~2010).
We can estimate when compressibility becomes important
by setting $|\u'| \sim H\Omega$. 
This gives a disruption time of
\begin{align}
t_2 \sim \frac{1}{s} \ln\left(\frac{H k_z}{a}\right).
\end{align}
Compressibility intervenes before the quadratic nonlinearities
only when $t_2< t_1$, which occurs if $H< \lambda_r$, a possible
regime for modes of moderate to small $k_z$. 
 
Alternatively, a runaway channel may be destroyed through the action of a
parasitic mode feeding off its strong shear and magnetic energy
(Goodman \& Xu 1994, Pessah \&
Goodman 2009, Latter et
al.~2009, 2010). By analogy with local boxes, we expect a variant of
the vertical Kelvin-Helmholtz instability to be the fastest growing
parasite. However, as argued in Appendix B in Latter et al.~(2010),
the orbital shear dramatically weakens the ability of the non-axisymmetric 
parasites to successfully destroy a channel mode: usually, a channel
is safe to grow to equipartition strengths. In cylindrical geometry,
it is less clear whether
parasitic modes are equally ineffective.

\section{Numerical simulations}

We now present a series of numerical simulations that
illustrate some of the properties discussed above. The first
objective is to follow the growth of the normal modes calculated
in Section~\ref{sec:linear_stability}, and then study the breakdown of
the linear regime. We start by describing the numerical method and the
setup of our simulations.

\subsection{Method and setup}

We use the finite volume code RAMSES (Teyssier 2002,
Fromang et al.~2006) to solve the axisymmetric 
\emph{compressible} MHD equations in a 2D
cylindrical coordinate system $(r,z)$.
Compressibility remains small during the linear phase
of the mode evolution, which RAMSES accurately captures
at the cost of a smaller
timestep. In addition, finite volume codes are now widely used in 
studying the properties and consequences of the MRI in astrophysical
disks; our use of a similar code will thus ease connections with
previously published results and aid future work.

 The
strategy we apply is the following: take a disk model in dynamical
equilibrium, superpose the eigenmodes calculated in Section 2.4.1
with a very
small amplitude, follow their growth and compute the associated growth
rate. We take the simplest possible initial disk state: 
an ideal gas of uniform density, uniform temperature (or,
equivalently, speed of sound $c_0$), and adiabatic index of $5/3$,
in Keplerian rotation around a
central point mass $M$. Units are fixed by choosing $GM=1$, while the
location of the grid's inner boundary is at $r=r_0=1$. We use $c_0=0.1$
though the remainder of this section.

Even though the setup is fairly  standard, two 
aspects deserve a more detailed discussion: the nature of the
thermodynamic perturbations
and the boundary conditions. 
 The
eigenmodes described earlier are in the incompressible limit and
involve a non-zero pressure perturbation 
(so that the velocity divergence associated
with the eigenmode remains zero). 
However, in compressible simulations, 
associated density perturbation will
feed into 
the momentum equation and create additional, non modal velocities
perturbations, complicating the interpretation of our
results. To avoid that problem, we introduce
temperature perturbations
that permit non-zero pressure perturbations but which keep
the density constant and uniform.

\begin{figure}
\scalebox{0.55}{\includegraphics{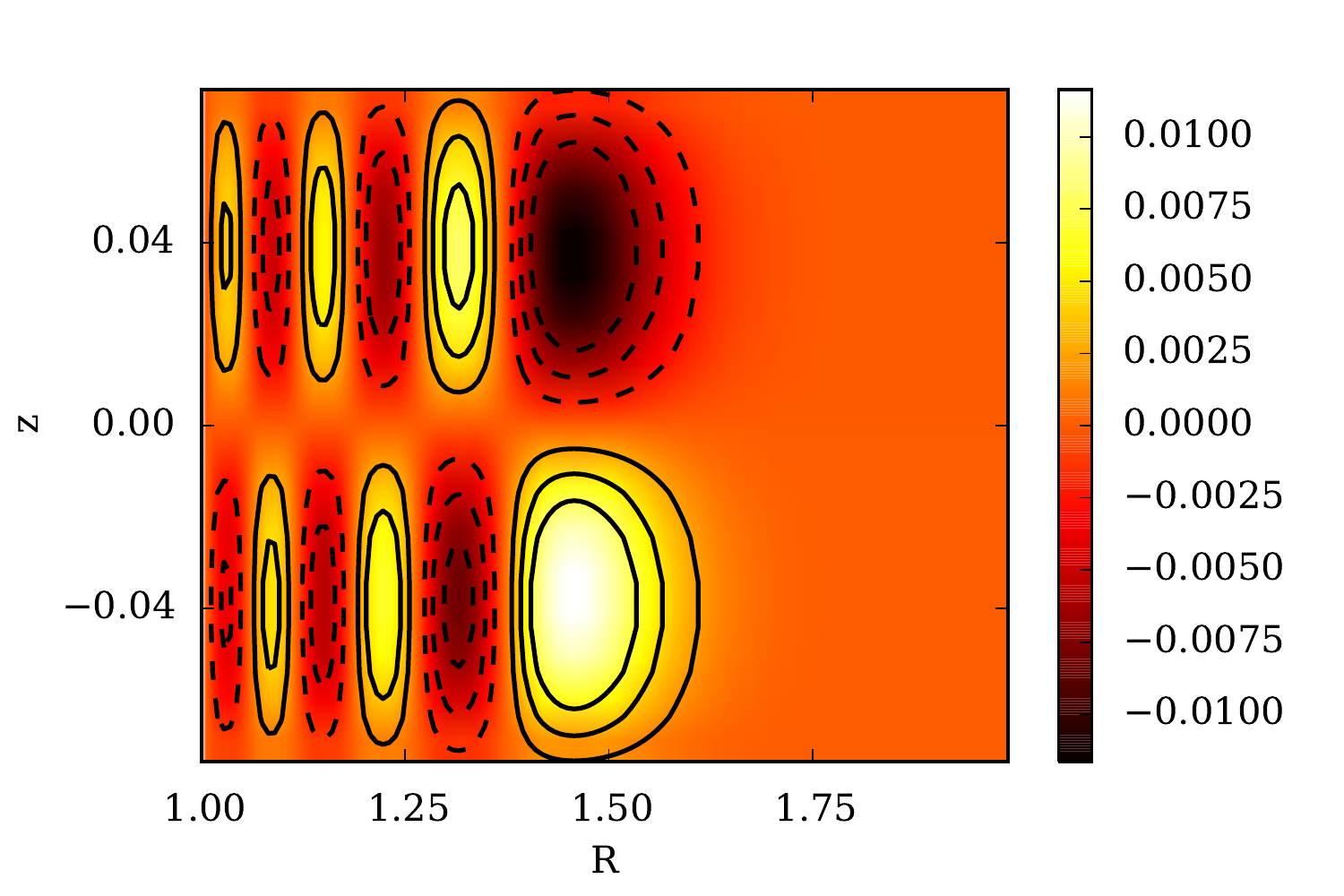}}
 \caption{Color contours showing the spatial distribution in the
   $(r,z)$ plane of $B_r/B_0$ for the mode with $k_z=40$ and $n=5$ at
   time $t=2$. The contour lines overplot $B_r/B_0$ at time $t=0$ and
   show that the mode has grown unperturbed over that period. Contours
   are drawn from $-6 \times 10^{-5}$ to $+6 \times 10^{-5}$ every $2 \times
   10^{-5}$. Negative contours are dashed and the zero contour is
   omitted.}
\label{fig:br_k40n5_beta200}
\end{figure}

\begin{figure*}
\scalebox{0.4}{\includegraphics{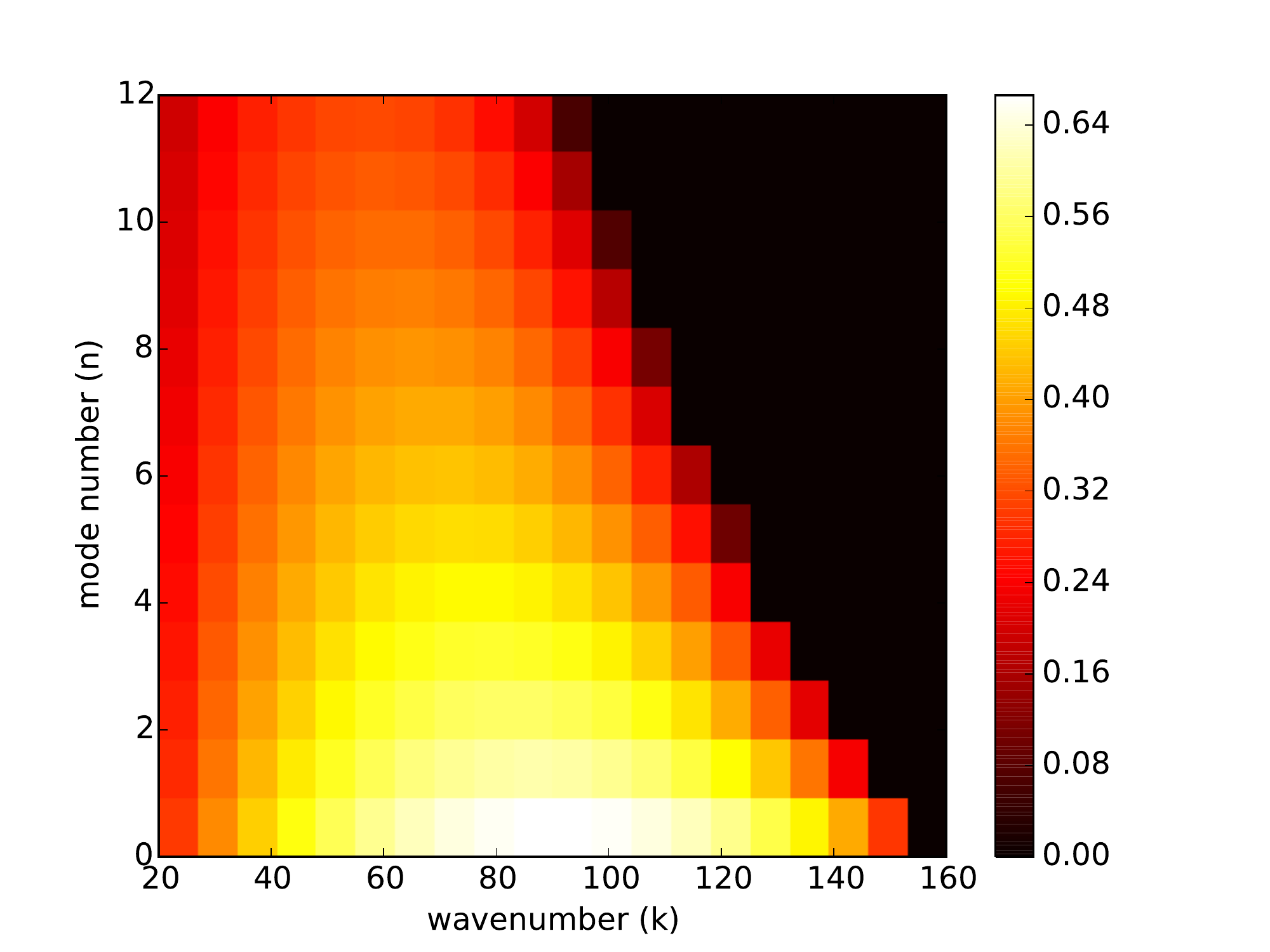}}
\scalebox{0.4}{\includegraphics{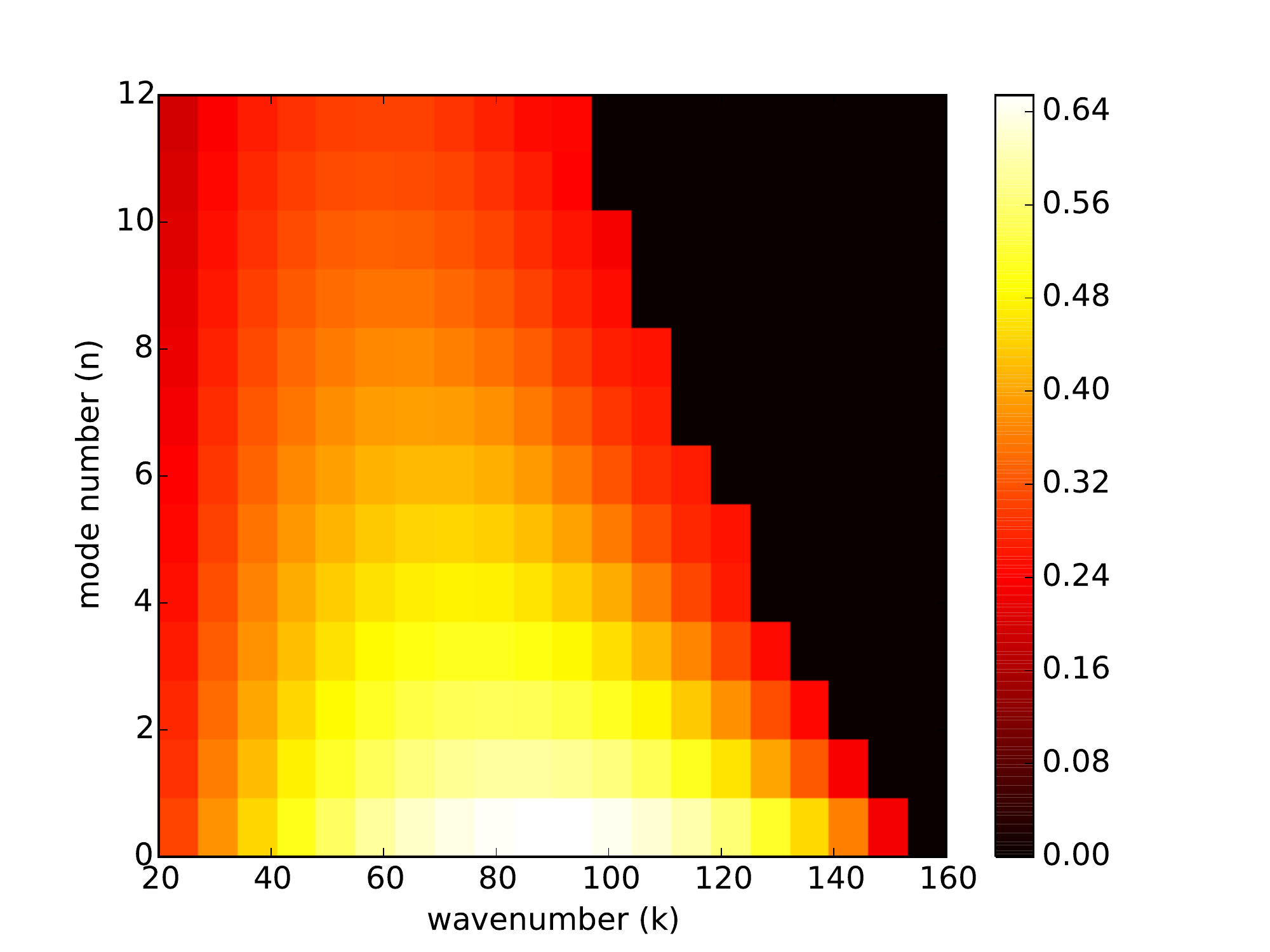}}
 \caption{Growth rates of the unstable modes in the $(k_z,n)$ plane for
   the case $\beta=200$ and $c_0=0.1$ determined using the linear
   analysis described in Section~\ref{sec:linear_stability} ({\it left
   panel}) and using an ensemble of $240$ numerical simulations ({\it
   left panel}) that have a common spatial resolution
   $(N_r,N_z)=(256,16)$.} 
\label{fig:parameter_survey}
\end{figure*}

The boundary conditions we used are periodic in the vertical
direction. This is possible in our setup because we neglect the
density's vertical stratification (see Section~\ref{sec:gov_eq}). 
The radial
boundary conditions are more subtle to implement. As discussed earlier,
the eigenmodes are confined to the inner disk. Thus, at the
outer radial boundary, we force the variables to take their
equilibrium, unperturbed values. At the inner boundary, we decompose
the variables in the ghost cells into the sum of their equilibrium
values (known analytically) and a perturbation. The latter are
chosen using the first active cells of the grid with the constraint
that they should satify the symmetry of the eigenmodes: 
zero-gradient for the vertical velocity and magnetic field perturbations
and anti-symmetric for the radial and azimuthal components of the
velocity and magnetic field (of course, the density perturbation
vanishes). With these boundary conditions, the disk equilibrium is
conserved to within machine accuracy and, as we shall see below, we
can follow the MRI eigenmodes to saturation.

\subsection{Linear growth}

In this section, we numerically compute the growth rate of the normal
modes for a vertical magnetic field whose strength is such
that $\beta$, the ratio between thermal and magnetic pressure, is
equal to $200$ (or, equivalently, $M_A=100$ given our value of the
sound speed). For a given eigenmode (defined by the values of $k_z$
and $n$), we compute the mode's spatial structure and theoretical growth
rate as described in Section~\ref{sec:num_eigen}. At $t=0$, we add the
perturbation associated with that mode to the equilibrium disk
structure and start the simulation. Its amplitude is such that the
maximum perturbed radial magnetic field amounts to $10^{-4}B_0$. The
computational domain extents radially out to $r_1=3$ and the vertical
size of the domain is set equal to the vertical wavelength of the
mode. The value of $r_1$ is here chosen large enough to ensure
that all variables reach their equilibrium value (i.e the mode amplitude
goes to zero) well within the domain. In order to evaluate the
amplitude of the mode during a simulation, we compute the time
evolution of the volume averaged rms of the radial component of the
magnetic field $B_r$: 
\begin{equation}
{\cal L}_2(B_r)=\left( \frac{\int\!\!\!\int r B_r^2 dr dz}{\int\!\!\!\int r dr dz} \right)^{1/2}
\end{equation}
The instantaneous growth rate $\sigma_{n}$ is then defined as the time
derivative of $\log({\cal L}_2)$. We denote by $\overline{\sigma_{n}}$ its
value averaged in time over the first two orbits of the simulation
(this short timescale ensures that the system remains 
within the linear phase).

To illustrate our general results we consider the specific
 mode $k_z=40$ and
$n=5$. Its theoretical growth rate is $\sigma=0.384$. After simulating
the mode's evolution for
 various grid sizes, we found that a
resolution of $(N_r,N_z)=(256,16)$ is required to capture correctly the
growth rate. In this case,
$\overline{\sigma_{n}}=0.380$, which corresponds to the theoretical
growth rate to better than a percent. In addition, the mode structure
is not modified during the linear growth. This is illustrated in
Fig.~\ref{fig:br_k40n5_beta200}, which shows a good correspondance
between the mode spatial structure in the $(r,z)$ plane at $t=0$ 
(contour lines) and $t=2$ (color contours), despite an amplification
by about three orders of magnitude. We note that the relative density
fluctuations remain smaller than $10^{-4}$ at $t=2$, which explains
the good agreement between the compressible simulation and the 
incompressible linear analysis. Decreasing the number of vertical cells
to 8 gives 
$\overline{\sigma_{n}}=0.339$, and the vertical structure of 
the mode is incorrectly described. 
Likewise, when decreasing the radial resolution by a
factor of two, so that $N_r=128$, the mode structure in
the inner parts of the disk is modified, even though 
$\overline{\sigma_{n}}=0.371$, which is a reasonable estimate. 
The mode's radial wavelength decreases inward and requires a fine resolution
to be properly captured.

Based on these results, we kept a fixed resolution of 
$(N_r,N_z)=(256,16)$ and systematically evaluated
$\overline{\sigma_n}$ for an ensemble of $240$ simulations in which we
varied the vertical wavenumber $k_z$ between $20$ and $160$ and the mode
number $n$ between $0$ and $12$. The results are summarized in
Fig.~\ref{fig:parameter_survey} and show excellent agreement
between the theoretical expectations ({\it left panel}) and growth
rates estimated from the numerical simulations ({\it right
panel}). Discrepancies occur for the slowest growing modes because
they are typically overwhelmed by faster growing modes
seeded by truncation errors. 
Except for these cases, we conclude that 
a finite volume
compressible code can
reproduce the results of the incompressible linear analysis.

\subsection{Nonlinear Saturation}

\begin{figure}
\scalebox{0.4}{\includegraphics{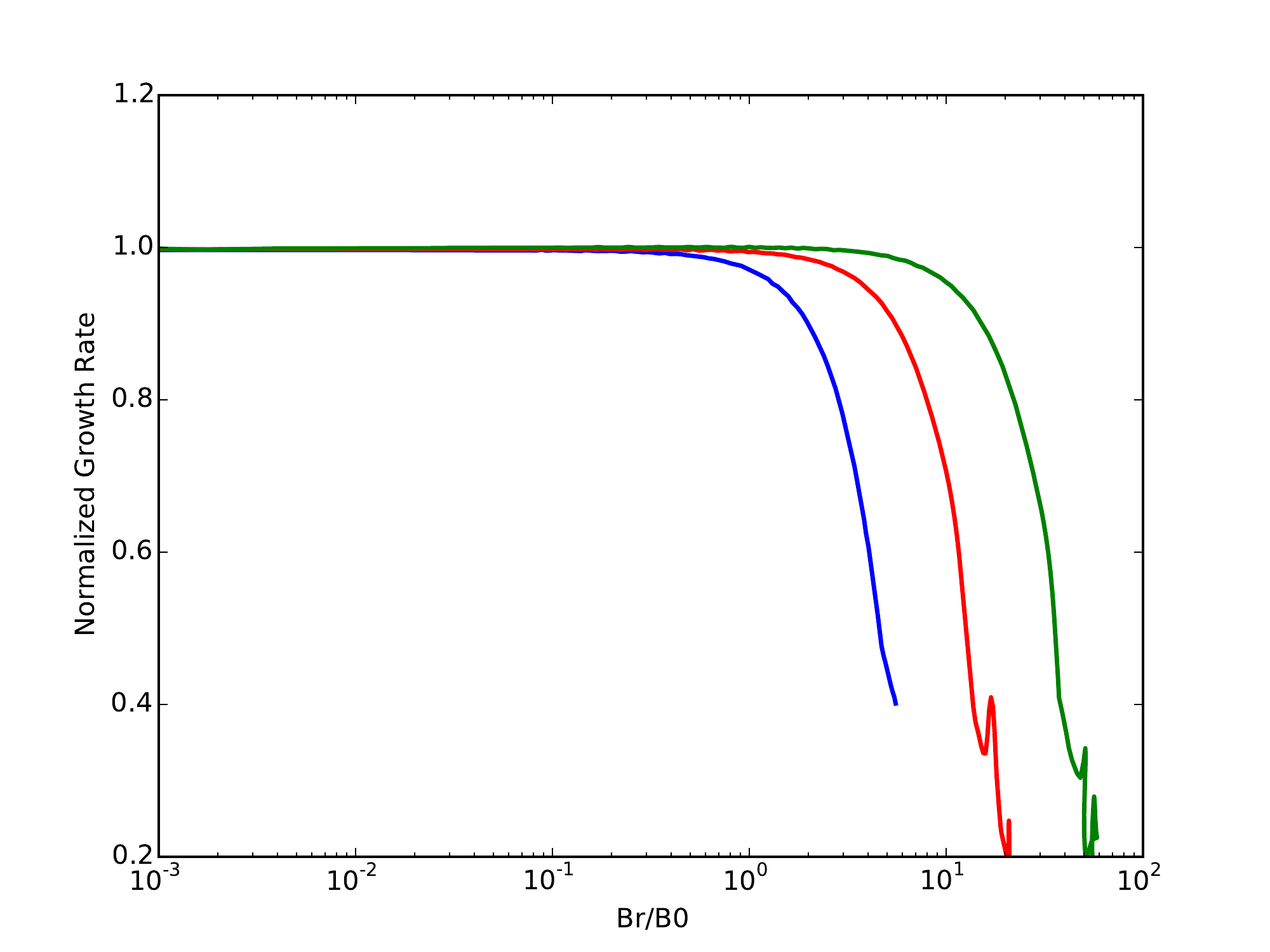}}
 \caption{Instantaneous growth rates normalised by the theoretical
   growth rate as functions of $B_r/B_0$, the maximum value
   of the perturbed radial field. In each of the three cases $n=0$, but
   $k_z=90$ ({\it blue curve}), $k_z=900$ ({\it red curve}), and
   $k_z=9000$ ({\it green
     curve}). The corresponding plasma betas are $\beta=200,\,2\times
   10^4$, and $2\times 10^6$.}
\label{fig:nonlinear_betavar}
\end{figure}

We now determine what happens when the modes 
enter the nonlinear regime. The simulations remain axisymmetric,
even though realistically the final saturated state will be
non-axisymmetric. 
Our aim, however, is not to
characterize the turbulence that follows the
linear mode breakdown, a formidable task that is well beyond
the scope of this paper. Rather, we ask the simpler
questions: what is the amplitude of a mode when it stops growing
exponentially with time? Do there exist linear modes that penetrate the nonlinear
regime while maintaining their structure, as predicted in Section 3? 

We performed three simulations for which $\beta=200$, $2 \times 10^{4}$ and $2
\times 10^{6}$. We focused on $n=0$ and select in each case
a mode with growth rate close to the maximum rate, thus
$k_z=90$, $900$ and $9000$,
respectively. Note that only in the last case is $\delta$ unambigously
small, being $\sim 0.05$ according to Eq.~\eqref{epscale}.  
The vertical and radial extent of
the computational domain is changed in each of the three simulations
to best accomodate each mode, as described
in the previous section. 

 Figure \ref{fig:nonlinear_betavar}
displays the instantaneous growth
rates $\sigma_n$ normalized by the theoretical growth rates $\sigma$ as a
function of $B_r/B_0$, where $B_r$ denotes here 
the maximum of the radial magnetic field fluctuation. 
Because $B_r/B_0$
increases monotonically it may be regarded as a proxy for time.
The nonlinear regime corresponds to $B_r/B_0>1$.
The figure clearly shows that the larger $k_z$,
the greater the amplitude achieved by the mode before
it breaks down. This is consistent with the prediction of Section 3,
which states that quadratic nonlinearities intervene only once a mode
grows to $\delta^{-1}$ the background. For the case $k_z=9000$, the
estimated maximum amplitude of the perturbed field is 20, in
fair agreement with Fig.~8, which shows that $\sigma_n/\sigma$ reaches
$0.9$ at this point. This mode, in particular,     
has penetrated significantly into the nonlinear regime.
This is not so convincingly the case for the lower $k_z$ modes, as
expected.

\begin{figure*}
\scalebox{0.5}{\includegraphics{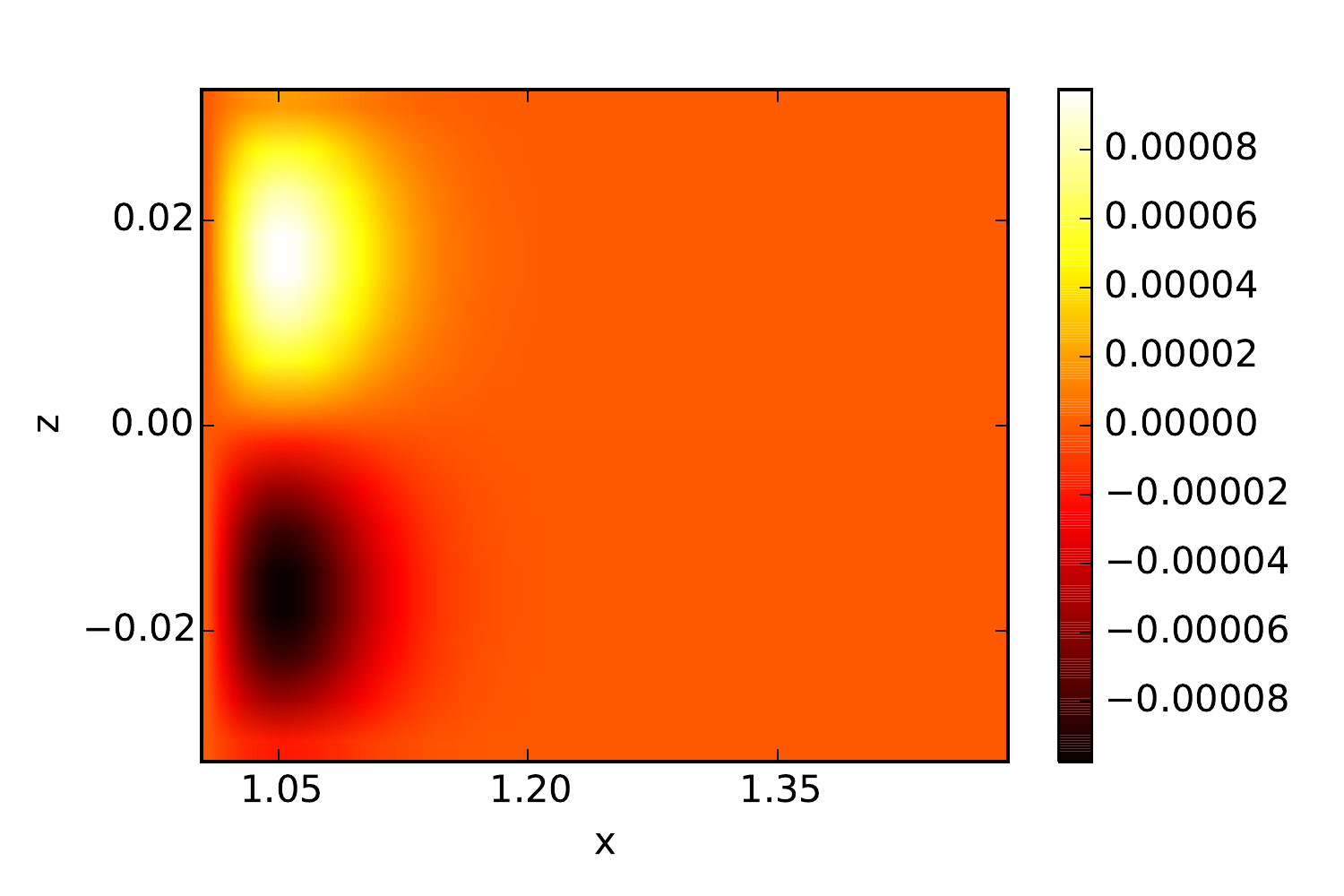}}
\scalebox{0.5}{\includegraphics{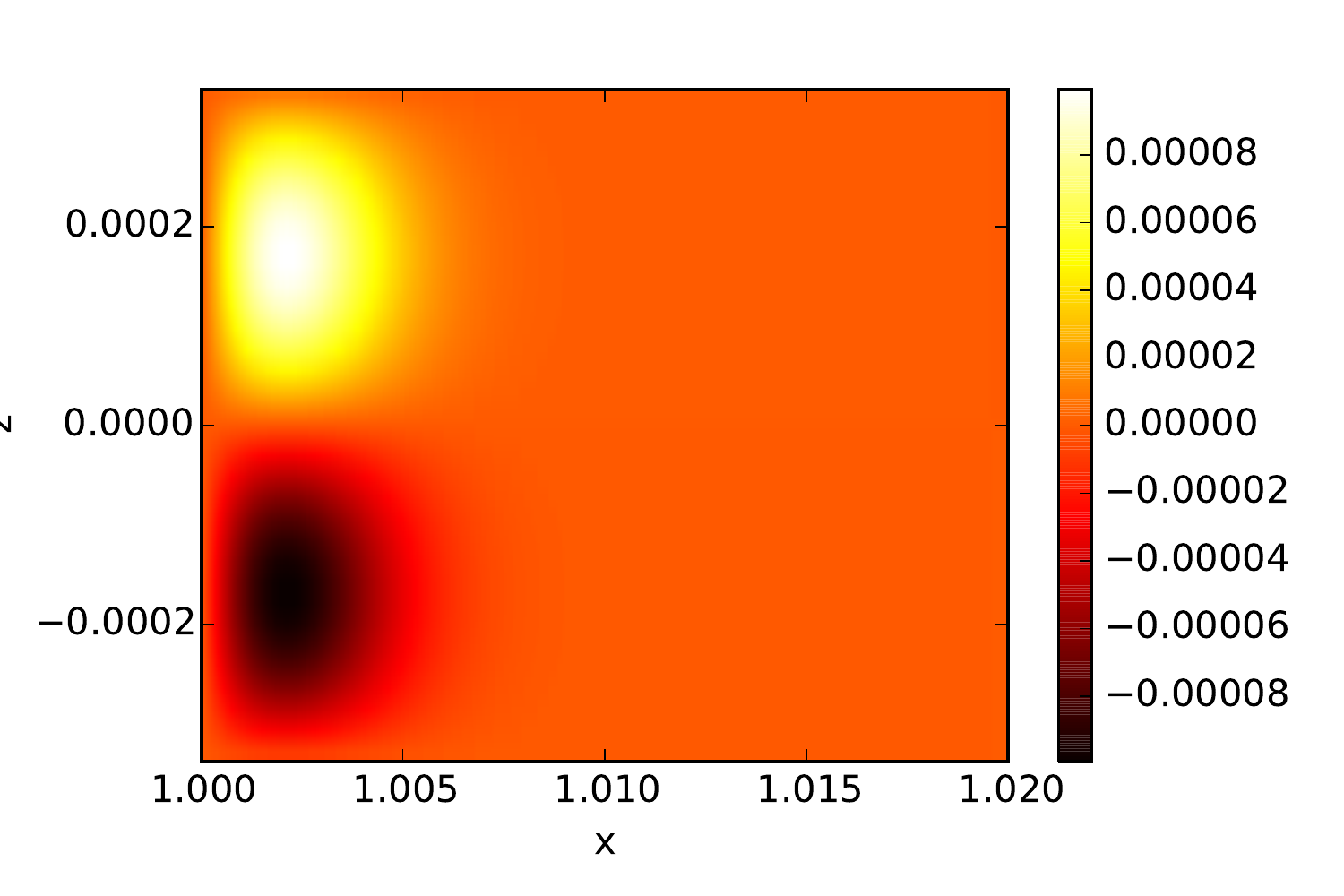}}
\scalebox{0.5}{\includegraphics{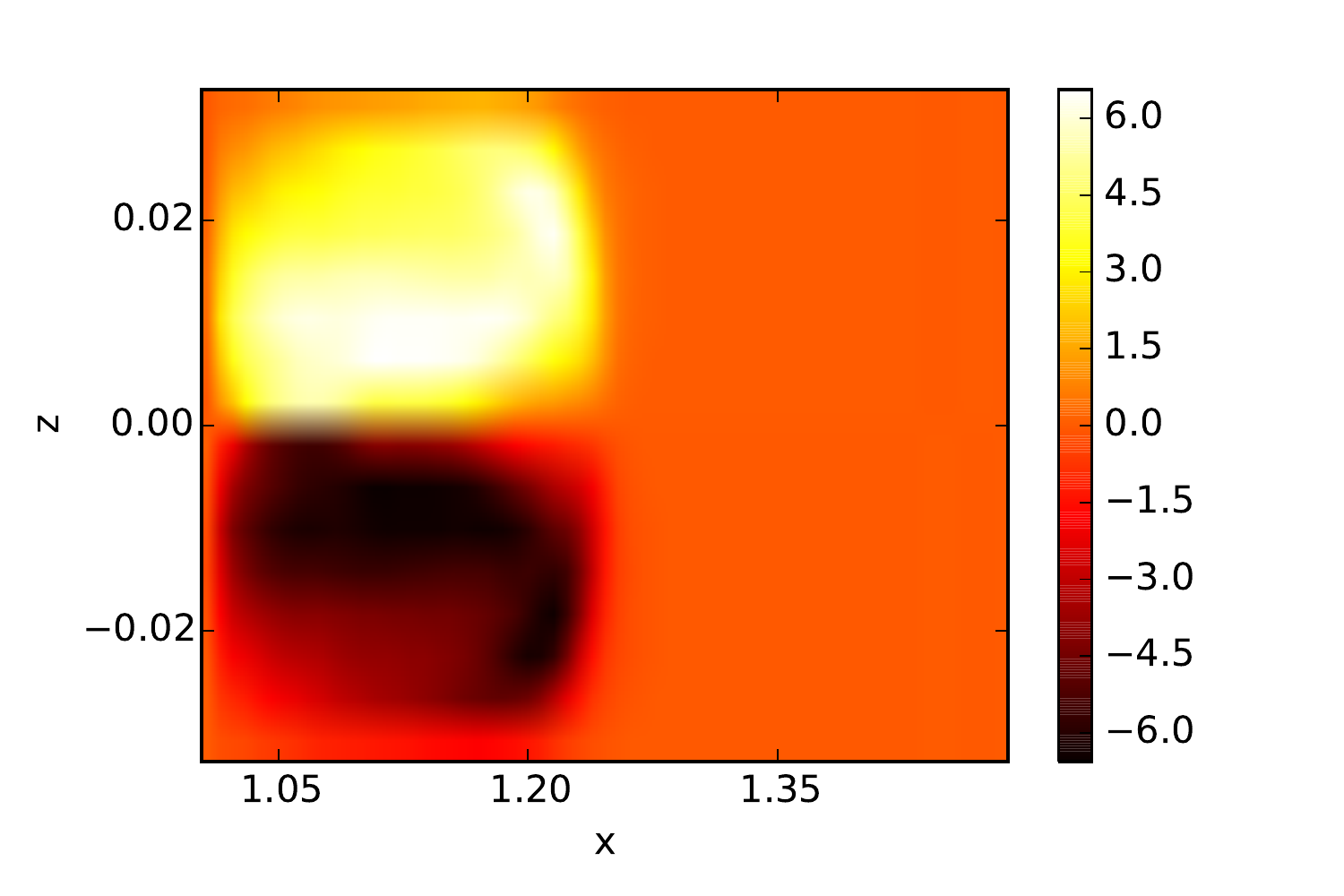}}
\scalebox{0.5}{\includegraphics{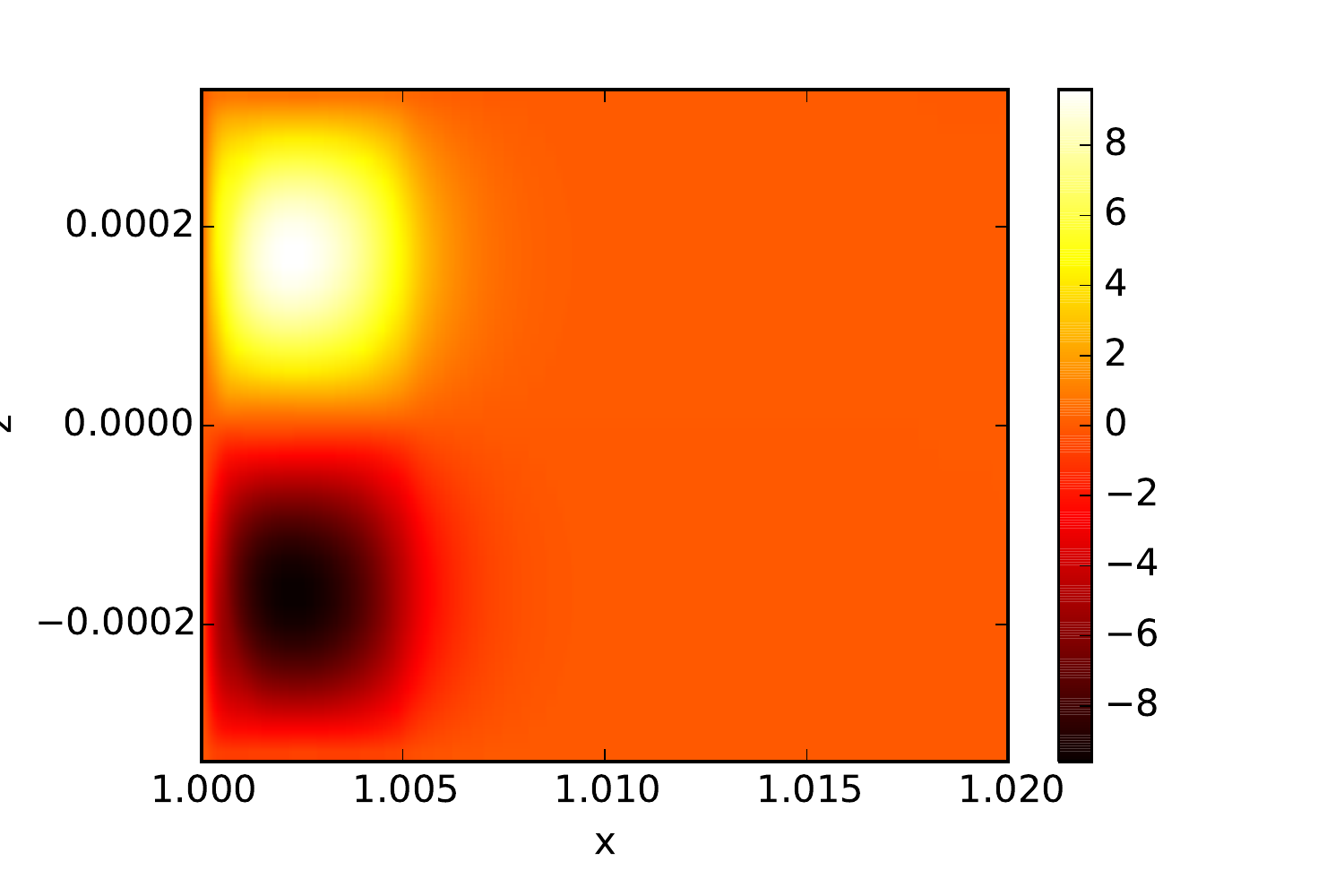}}
 \caption{Color contours of $B_r/B_0$ for $k_z=90$, $\beta=200$ ({\it left
     column}) and $k_z=9000$, $\beta=2 \times 10^6$ ({\it right column}) at $t=0$
   ({\it top row}) and at $t=3.2$ ({\it bottom left panel}) and $2.5$
   orbits ({\it bottom right panel}). By $t=2.5$, the
  $k_z=9000$ mode has grown by
 five orders of magnitude and is ten times larger than the background
 field, and yet it has retained its linear structure to a good
 approximation despite being well inside the nonlinear regime.}
\label{fig:BrB0VarBeta}
\end{figure*}

In Fig.~\ref{fig:BrB0VarBeta} we compare the
distribution of the radial magnetic field at $t=0$ ({\it top row})
and at a later time when $B_r$ has grown to amplitudes larger than
the background vertical field ({\it bottom row}) for $k_z=90$ and
$k_z=9000$. 
The figure shows that the low $k_z$ mode 
has become strongly disturbed by
$t=3.2$, whereas the field of its large $k_z$ counterpart
retains a structure close to linear, despite possessing an
amplitude 10 times greater than the background.

Finally, we check the role of compressibility, which should be
important when $H<\lambda_r$. In code units, $H=0.1$ at $r=r_0$, whereas
$\lambda_r\sim 0.1$ and $\lambda_r\sim 0.001$ for the two modes $k_z=90$ and
$k_z=9000$. Compressibility is certainly subdominant in the
breakdown of the large $k_z$ mode, which we
attribute to quadratic nonlinearities. In the breakdown of the 
low $k_z$, on the other hand, compressibility and nonlinearity are of
equal importance.  
 
\section{Conclusion}

We have presented a linear analysis of the MRI in
cylindrical geometry in order to make clear the connection
between the global and the local theories. 
In particular, we show that
each local channel mode corresponds to the evanescent part of a global
mode. Local radially varying modes, on the other hand, 
correspond to sections of the same
global mode at smaller radii. Moreover, we show that only global
axisymmetric modes of extremely large vertical wavenumber $k_z$
are approximate nonlinear solutions to the governing
equations. As these modes are localised to the inner boundary, most of
the disk never experiences the `nonlinear property' of the MRI.
Direct simulations with RAMSES verify
the last point, which also provides a useful numerical check on global
codes generally. Our results raise a number of questions and issues,
which we now list.

First, one of the most notable features of channel flows in local
boxes is that they are nonlinear solutions to the governing equations.
But as this property fails to appear in global disk models it is
clear that this feature is an artefact of the local approximation.
A natural question is then: how seriously does this artefact distort
the nonlinear dynamics of the MRI in shearing boxes?   
It is well known that channel flows strongly influence
the MRI saturation in certain circumstances (e.g.\ Sano \&
Inutsuka.~2001, Bodo et
al.~2008, Lesaffre et al.~2009, Murphy \& Pessah 2015). 
How representative of global MRI turbulence is
local MRI turbulence, given
the artifical prominence of channel flows in the latter?

Second, how non-local is the onset of the MRI and its ensuing
turbulence in global disks? 
At a given radius $r_*$, the fastest growing disturbance is the
 normal mode that has its
turning point there. But such a mode 
 extends from the inner boundary $r=1$ to $r=r_*$, 
so its growth, and subsequent behaviour,
may be influenced by
all the shorter time-scale activity on the intervening 
shorter radii. Indeed, because
different global modes encompass overlapping regions, 
significant mode-mode interactions should arise: modes
 that extend over large radii should interact with faster growing modes
 localised to small radii.
 Fluctuations at outer
 radii may be `slaved' to what is going on at smaller radii. Of
 course,
distant regions must decouple on some scale, 
 but what sets that scale? Also how might
 all this be
 connected to global field generation and dynamo action? Such
 questions can only be answered by careful numerical simulations. But
 an understanding of the normal mode structure may provide useful
 clues.

Generalising the equilibrium magnetic field 
is another avenue to explore. Neglected in this paper, azimuthal
fields are probably the dominant component in a differentially
rotating flow; their importance in the disk's stability could be
reassessed in an analogous way to here. In addition, a radially
varying magnetic equilibrium should also be revisited. Crucially, a
vertical field that
decays with radius will alter the locations of the modes' turning
points; modes of given $k_z$ and $n$ will extend further outward,
changing the nature of the disk's linear response.

A fourth issue regards the role of large-scale
non-axisymmetric MRI modes, something we have not touched on. Such modes
must contend with magnetic resonances and consequently,
their localisation is more
complicated than for the axisymmetric MRI (e.g.\ Curry \& Pudritz
1996, Fu \& Dong 2009).
How does this influence, at all, their participation in disk
turbulence?

Other topics of interest include
the breakdown of the linear modes due to non-axisymmetric parasitic
instabilities (Goodman \& Xu 1994), 
which may provide an alternative pathway to saturation
in some circumstances. Three-dimensional cylindrical simulations could
probe their behaviour and assess their relative importance. 
A number of weakly non-linear analyses of the MRI have
been conducted, all using local approximations (Umurhan et
al.~2007, Jamroz et al.~2008, Vasil 2015). 
It would be interesting to test if
analogous calculations are possible in a cylindrical model, using the
formalism presented in this paper as the (linear) starting point.
Finally, 
our cylindrical results could
be extended to vertically structured global disk
models, obviously making contact with the general theory  
of Terquem \& Papaloizou (1996) and Ogilvie (1998), 
but also exploring instabilities in the
strong magnetic field limit 
(cf.\ Curry \& Pudritz 1995, Pessah \&
Psaltis 2005). The latter may be of particular interest to
the magnetically arrested accretion flows around black holes 
recently simulated 
(e.g.\ Tchekhovskoy et al.~2011, McKinney et al.~2012).

\section*{Acknowledgements}
The authors thank the anonymous reviewer for a very prompt and helpful
set of comments.
HNL is partially funded by STFC grant ST/L000636/1. 
SF acknowledges funding from the European Research Council under the
European Union's Seventh Framework Programme (FP7/2007-2013) / ERC
Grant agreement n258729.

\begin{appendix}

\section{Mathematical derivations}

\subsection{Eigenvalue ordering}

In this short section we sketch out a proof showing that the
eigenvalues $\varepsilon$ associated with
\eqref{Und} are always less than 1. For simplicity the boundary
conditions are taken to be either $u_r'=0$ or $\d_r u_r'=0$ at
$r=r_1,\,r_2$. The proof for free boundaries is a little more involved
and we omit its details.

First multiply
\eqref{Und} by $U^*$ and integrate over the domain. After integrating
by parts and applying the boundary conditions the equation can be
reworked into
\begin{align}\label{A1}
\varepsilon^{-2} = \frac{\int |U|^2 dr}{\int r^{-2q}|U|^2 dr} + \frac{\int
  |\d_rU|^2 + \tfrac{3}{4}r^2|U|^2\,dr}{k_z^2\int r^{-2q} |U|^2 dr}.
\end{align}
We see straightaway that $\varepsilon$ must be positive. But note also
that $|U|^2 > r^{-2q}|U|^2$ over the entire integration range and so the
first term in Eq.~\eqref{A1} must be greater than 1. As a consequence,
$ \varepsilon < 1$.

\subsection{Eigenvalues in the large $k_z$ limit when $q=1$}

Here we obtain approximate solutions to the eigenvalue equation
$K_\nu(k_z)=0$ in Section 2.4.2. In the limit of large $k_z$ both the
order and argument of the Bessel function go to infinity. We call on
the asymptotic expression given in Ferreira \& Sesma (2008) for the
roots of $K_\nu(x)$ when the order $\nu$ is large and imaginary:
\begin{equation}
x_n \approx \text{e}^{-\text{i}\pi/2}(\nu -2^{-1/3}a_n \text{e}^{-2\pi\text{i}/3}\nu^{1/3}).
\end{equation}
Here $a_n$ is the $n$'th root of the Airy function
$\text{Ai}(x)$. Substitution of $\nu= \text{i}k_z/\varepsilon$
obtains the cubic equation
\begin{equation}
\varepsilon -a_n\,2^{-1/3}k_z^{-2/3}\,\varepsilon^{2/3}-1 =0,
\end{equation} 
the correct root of which can be approximated explicitly by expanding
$\varepsilon$ in powers of small $k_z^{-2/3}$ around $1$.  

\end{appendix}
\end{document}